Quantized spin waves and perpendicular standing spin waves stimulated by current in a single-layered ferromagnetic wire

A. Yamaguchi[1,2], K. Motoi[1], and H. Miyajima[1]

[1] Department of Physics, Keio University, Hiyoshi, Yokohama 223-8522, Japan

[2] PRESTO, JST, Sanbancho 5, Chiyoda, Tokyo 102-0075, Japan

**[Abstract]**

The rectifying effect of radio-frequency (RF) current is highly sensitive in terms of the spatial spin distribution and dynamics. It emerged that an additional spin wave mode was stimulated by the direct-current (DC) current and that this spin wave was detectable via rectification of the RF current. A phenomenological model to describe the time-dependent anisotropic magnetoresistance or time-dependent planer Hall effect is proposed and found to correlate well to the experimental results. The nonlinear spin dynamics accompanying additional spin waves are studied as functions of the RF and DC currents, the external magnetic field, and the applied field direction.

The understanding of spin dynamics in artificial nano-magnets is important, not only in fundamental magnetism but also in the technological areas. One of the distinctive spin dynamics in the high frequency region is the spin wave excitation: an extended ringing of the magnetization produces high frequency resonance characteristic to the devices[1, 2]. In nano-scale ferromagnetic devices, both exchange and dipole energies contribute to the spin wave spectrum, which is strongly dependent on the system geometry[1, 2]. The spin wave resonance and magnetization dynamics in the confined geometry have been investigated via Brillouin light scattering (BLS)[2, 3], ferromagnetic resonance (FMR), time-resolved magneto-optical Kerr effect[4 - 7], the rectifying effect[8 – 13], and so on.

When the spin-polarized current traversing a ferromagnetic conductor transfers its spin-angular momentum to the magnetic system[14, 15], it causes the magnetization to precess with strong s-d exchange interaction between the conduction electron and the magnetic moment[14, 15]. Recently, electrical detections of the magnetization motion have been performed; not by applying an alternating-current (AC) magnetic field but by using the spin-transfer torque of a spin-polarized AC current[9 - 11]. It also reveals a detailed understanding of the spin dynamics due to the conduction electrons and the magnetic moments.

As is well known, the propagation of electromagnetic waves though the ferromagnetic conductors produces some nonlinear effects, reflecting the interaction between current and magnetization via magnetoresistance and extraordinary Hall effects. The non-linear effect occurs remarkably within the frequency region behavior of FMR and is detectable by using electrical measurements. We have studied the magnetic behavior of FMR in a single layered $Ni_{81}Fe_{19}$ ferromagnetic wire by the simultaneous application of direct-current (DC) and radio-frequency currents.

In a previous paper[11], we proposed a phenomenological model for the magnetoresistance response induced by the nonlinear effect and showed a consistent view describing the DC voltage generation in a single-layered ferromagnetic wire. In this paper, we propose an analytical model; taking advantage of the well-resolved frequency-domain DC spectra induced by the rectifying effect. Not only the quantized spin wave excitation but also the perpendicular standing spin wave (PSSW)[1] are induced by the DC current through the rectification of the PHE.

The experiments are performed on a 30nm-thick $Ni_{81}Fe_{19}$ wire. The wire is fabricated onto an MgO substrate via electron beam lithography and the liftoff method. Figure 1 shows an optical micrograph of the wire of width 5μm together with the electric measurement circuit. The sinusoidal constant wave (CW) RF current with current density of $3.0 \times 10^{10}$ A/m$^2$ is injected into the wire by a signal generator with a frequency range from 10 MHz to 15 GHz. Simultaneously, the DC current is applied to the wire through the bias-tee, which separates the

DC- and RF-components of the current and the external magnetic field $H_{ext}$ is applied in the substrate plane as a function of angle $\phi$ from the major axis of the wire. The precession of the magnetic moment in the vicinity of FMR region generates the DC voltage attributable to the magnetoresistance oscillation. The experiment is performed at room temperature with the slowly sweeping frequency of the RF current that flows along the major axis of the wire. The Hall voltage spectra, $V_{Hall}$, induced across the minor axis of the wire, is also measured.

On the standpoint of the Mott's two current model, the electrical field $E$ in a ferromagnetic metal is given in the following form:

$$E = \rho_\perp j + m(j \cdot m) \cdot (\rho_\parallel - \rho_\perp) + \rho_H m \times j, \qquad (1)$$

where $j$ denotes the electrical current density, $m$ the unit vector along the local magnetization, $\rho_\perp$ and $\rho_\parallel$ the resistivity parallel and perpendicular to $j$ with respect to $m$ and $\rho_H$ the extraordinary Hall resistivity. Juretschke[8] introduced a phenomenological oscillating component of the magnetization to Eq. (1), and pointed out the possibility of the generation of the DC voltage when the magnetization precession was caused by the RF field.

In general, the magnetization process in a ferromagnetic wire is well described by the Stoner-Wohlfarth single domain model. As shown in Fig. 1, when the magnetization unit vector is located at the origin, $m = (\sin\theta\cos\phi, \sin\theta\sin\phi, \cos\theta)$, and the electrical current flows along the longitudinal axis of the wire, $j = (j,0,0)$, the electrical field $E$ is given by:

$$\boldsymbol{E} = \begin{pmatrix} E_x \\ E_y \\ E_z \end{pmatrix} = j \begin{pmatrix} \rho_\perp + (\rho_\parallel - \rho_\perp) \sin^2\theta \cos^2\phi \\ (\rho_\parallel - \rho_\perp) \sin^2\theta \cos\phi \sin\phi + \rho_H \cos\theta \\ (\rho_\parallel - \rho_\perp) \sin\theta \cos\theta \cos\phi - \rho_H \sin\theta \sin\phi \end{pmatrix}. \qquad (2)$$

When the magnetization directs along the major axis (the magnetic easy axis), $E_x$ and $E_y$ along the magnetic easy axis (x-axis) and the minor-axis (y-axis) in Eq. (2) are respectively given by:

$$E_x = j\left(\rho_\perp + (\rho_\parallel - \rho_\perp)\cos^2\phi\right), \qquad (3)$$

$$E_y = j(\rho_\parallel - \rho_\perp)\cos\phi \sin\phi. \qquad (4)$$

These Eqs. (3) and (4) are known to represent the AMR effect and the planer Hall effect, respectively.

Taking into account the forced magnetization precession in the presence of a spin transfer effect[14, 15] and the non-uniform internal magnetic field[16], the electric field components $E_x$ and $E_y$ are given by:

$$E_x(t) = j\left\{\rho_\perp + (\rho_\parallel - \rho_\perp)\cos^2(\phi + \delta(t))\right\}, \qquad (5)$$

$$E_y(t) = j(\rho_\parallel - \rho_\perp)\cos(\phi + \delta(t))\sin(\phi + \delta(t)). \qquad (6)$$

By expanding the Eqs. (5) and (6) by the addition formula, we can approximately derive the average time-dependent electric fields due to the AMR effect and planer Hall effect as:

$$E_x(t) \approx j\left[\rho_\perp + (\rho_\parallel - \rho_\perp)\left(\cos^2\phi - \left\langle \frac{1}{2}\sin 2\phi \sin 2\delta(t)\right\rangle\right)\right], \qquad (7)$$

$$E_y(t) \approx \frac{1}{2} j\left[\sin 2\phi + \left\langle \cos 2\phi \sin 2\delta(t)\right\rangle\right]. \qquad (8)$$

When the CW RF current is introduced into the wire in the FMR state, the magnetization motion is a steady-state precession mode, compensating for the actual loss of the precession amplitude. Suppose that $j(t) = j_0 \cos(\omega t)$, the frequency variation of the induced voltage, as derived from the AMR effect and the PHE, is given by the Fourier transformation of $E(t)$;

$$E_x(\omega) = j_0^2 A(\omega) \sin 2\phi \cos\phi, \tag{9}$$

$$E_y(\omega) = j_0^2 A(\omega) \cos 2\phi \cos\phi, \tag{10}$$

where $A(\omega)$ is the frequency spectrum given in the previous papers[11, 17].

Figure 2(a) shows the RF frequency dependence of the output signal $V_{\text{Hall}}$ for the DC current $I_{\text{DC}}$ = +15 mA (red solid line) and $I_{\text{DC}}$ =0 mA (black solid line) in the external magnetic field $|H_{\text{ext}}|$ = 0.5 kOe applied at the angle $\phi = 45°$. Here, the non-resonant background, which is much stronger than the resonant signal, is subtracted for clarity. As shown in Fig. 2(a), the spectrum for $I_{\text{DC}} = 0$ contains at least two distinct modes near 6.2 GHz and 7.1 GHz, while the spectrum for $I_{\text{DC}} = +15\,\text{mA}$ has a additive distinct modes near 10.8 GHz. Figure 2(b) shows how the relationship between $I_{\text{DC}}$, and the Hall voltage difference $\Delta V_2$ can be well described by $\Delta V_2 = \Delta R_{\text{PHE}} \cdot I_{\text{DC}}$, where $\Delta R_{\text{PHE}}$ is the planer Hall resistance. The estimated $\Delta R_{\text{PHE}}$ is $\Delta R_{\text{PHE}}$ =0.032 mΩ for $(\phi, |H_{\text{ext}}|) = (45°, 0.5\,\text{kOe})$ and $\Delta R_{\text{PHE}} = -0.053$ mΩ for $(\phi, |H_{\text{ext}}|) = (120°, 0.2\,\text{kOe})$. It should be noted that the sign and value of $\Delta R_{\text{PHE}}$ strongly correlates with the direction of the magnetization and precessional angle.

The magnetic field dependence of the spin wave frequency of each spin mode is shown in Fig. 3 and all observed modes are confirmed to be attributable to the magnetic excitations. The two modes observed at a lower frequency (red circles and blue squares) and one mode at the higher frequency (black triangles) correspond to the quantized spin waves derived from the quantized Damon-Eshbach mode and the PSSW in a single wire, respectively[1,2,18]. We obtain an empirical expression describing the complete spin wave modes with the quantized integer numbers.

According to Kalinikos and Slavin[1,2,18], the dispersion of spin waves in a confined magnetic structure is given by:

$$\omega^2 = \gamma^2 \left( H + \frac{2A}{M_S} q^2 \right) \left( H + \frac{2A}{M_S} q^2 + 4\pi M_S \cdot F_{pp}(q_\| d) \right), \tag{11}$$

where

$$q^2 = q_x^2 + q_y^2 + \left( \frac{(p - \Delta p)\pi}{d} \right)^2 = q_\|^2 + q_\perp^2, \tag{12}$$

for the normal to the film surface, $q_\|$ is the in-plane wave vector. Here, since the wire length is much longer than the width, the magnetization vector almost aligns parallel to the longitudinal axis of the wire (x direction). Suppose that the quantization along the x direction is neglected and then the y component is quantized as $q_\| = n\pi/w$, where $w$ is the wire width and the integer $n$ is the quantization number for the quantized Damon-Eshbach mode (Q-DE). The quantization of the $z$ component is given by $q_\perp = (p - \Delta p)\pi/d$ for the wire thickness $d$, while the quantized number $p$ is the integer of a half wavelength along the $z$ direction, which corresponds to the PSSW. Meanwhile, the correction factor $\Delta p$ ( $0 \leq \Delta p \leq 1$ ) is determined by the boundary condition, and $F_{pp}(q_\| d)$ is the matrix element of the magnetic dipole interaction[17]. In the present analytical spin wave modes, these parameters are determined as $(n, p-\Delta p)$=(0, 0), (1. 0) and (0, 0.47) and the calculated lines using them correlate well to the experimental data. As shown in Fig. 3, the size effect is easily taken into account for the long wire in a single domain state because the dipole field is homogeneous, whereupon we obtain the so-called Kittel mode[1, 19]:

$$\omega^2 = \gamma^2 \left(H + \left(N_y - N_x\right)M_S\right) \cdot \left(H + \left(N_z - N_x\right)M_S\right), \qquad (13)$$

where $M_S$ denotes the saturation magnetization, and $N_x, N_y, N_z$ the demagnetization coefficient along the x, y, and z axes. In the present case of the line shape geometry, $N_x \approx 0$ and $N_z \approx 1 - N_y$, and the dynamic demagnetization field is assumed to be considerably homogeneous. The magnetic field dependence, as calculated by Eq. (13), is shown by the red dashed line in Fig. 3. The correlation with the experiment is convincing, indicating that the simple analysis is available for the rectification. Furthermore, it means that the RF current results in the resonant excitation of the pure uniform magnetization precession and that the current can excite various spin wave modes characteristic to the confined structure.

The significance of this experiment is that not only the Q-DE and PSSW modes are involved but also the DC current-induced PSSW mode. Here, we discuss the origin of the PSSW mode induced by the DC current. To simplify the analysis of the magnetization dynamics, we focus on the resonant excitation derived by the single mode of the magnetization precession, as described by Eqs. (8) and (10). Figure 4 shows the variation of the DC voltage difference $\Delta V$ with the applied magnetic field angle $\phi$; (a) $\Delta V_1$ for $I_{DC} = 0\,\text{mA}$, (c) $\Delta V_2$ for $I_{DC} = +15\,\text{mA}$, and (b) the non-resonant background voltage $V_{bg}$ for $I_{DC} = +15\,\text{mA}$. According to Eqs. (8) and (10), $V_{bg}$ and $\Delta V_1$ vary with the angle as $\sin 2\phi$ and $\cos 2\phi \cos \phi$, respectively. The voltage $V_{bg}$ is derived from the time independent term $\sin 2\phi$ in Eq. (8), while $\Delta V_1$ originates from the time dependent magnetization precession term in Eq. (8). Indeed, the experiment fits well to the present analytical curve, suggesting that the magnetization precession at the finite frequency is the uniform mode. Coversely, the angle dependence of $\Delta V_2$ with $I_{DC} = +15\,\text{mA}$ correlates well to the $\sin 2\phi$ line. This result indicates that the PSSW mode induced by the DC current may be related to the magnetization dynamics in the plane caused by the dynamic torque perpendicular to the film surface. There are two possibilities; one is the gradient of the magnetic field generated by the DC current flowing though the wire and electrodes[11], the other is the presence of the absorption and dissipation of spin angular momentum through the anisotropic energy $K_\perp$ (see Fig. 4(d), Tatara and Kohno[20]) or the spin relaxation (Zhang and Li[21]) derived from the spatial spin distribtuion[16]. In particularly, there is definite potential for the latter to play an important role in the spin dynamics in the RF region.

We showed the PHE rectification effect of the (DC+RF) current in a single layered ferromagnetic wire, which exhibits time-dependent magnetoresistance and reveals the PSSW induced by the DC current. Although the origin of the DC current-induced PSSW remains unclear at present, the results will provide a new way for studying spin dynamics in the confined magnetic structure with the correlation between the localized magnetic moments and the conduction electrons.

The present work was partly supported by the MEXT Grants-in-Aid for Scientific Research in Priority Area, the JSPS Grants-in-Aid for Scientific Research and the Keio Leading-edge Laboratory of Science and Technology project 2007.

[Figure captions]

**Figure 1**

Schematic diagram of RF measurement; the top view of the optical micrograph of the sample, and the model geometry and symbol definitions.

**Figure 2**

(a) Typical Hall voltage spectra with $I_{DC}=0\,\text{mA}$ and $I_{DC}=+15\,\text{mA}$ measured in the application of magnetic field $|H_{ext}|=0.5\,\text{kOe}$ at $\phi=45°$. Each spectrum is vertically shifted for clarity. (b) The DC current dependence of the Hall voltage difference $\Delta V_2$.

**Figure 3**

Resonant frequencies generated in the Hall voltage spectra as a function of the applied field. The lines are calculated using either a Q-DE equation with quantized wave vectors (black solid line labeled Q-DE), the Kittel formula (red dashed line), or the first PSSW mode (black solid line labeled PSSW). The typical Hall voltage spectrum is shown in the inset.

**Figure 4**

Magnetic field angle dependence of (a) the Hall voltage difference $\Delta V_1$ with $I_{DC}=0\,\text{mA}$, (b) the non-resonant Hall voltage $V_{bg}$ with $I_{DC}=+15\,\text{mA}$, and (c) the Hall voltage difference induced by the DC current $I_{DC}=+15\,\text{mA}$ in the magnetic field of $|\mathbf{H}_{ext}|=0.5\,\text{kOe}$.

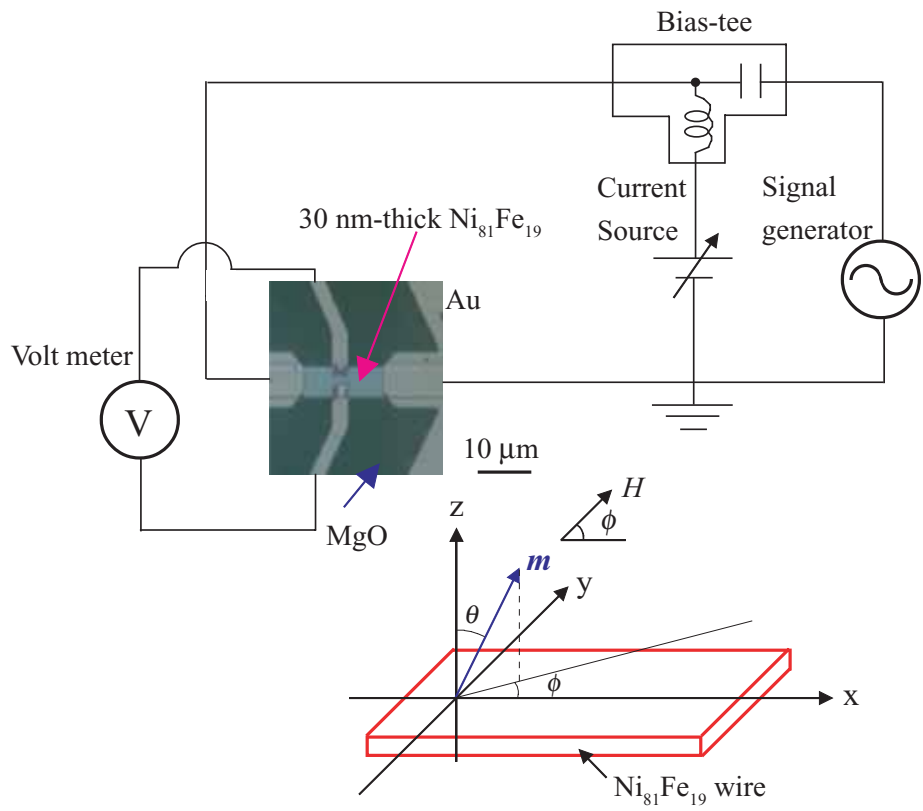

Fig. 1

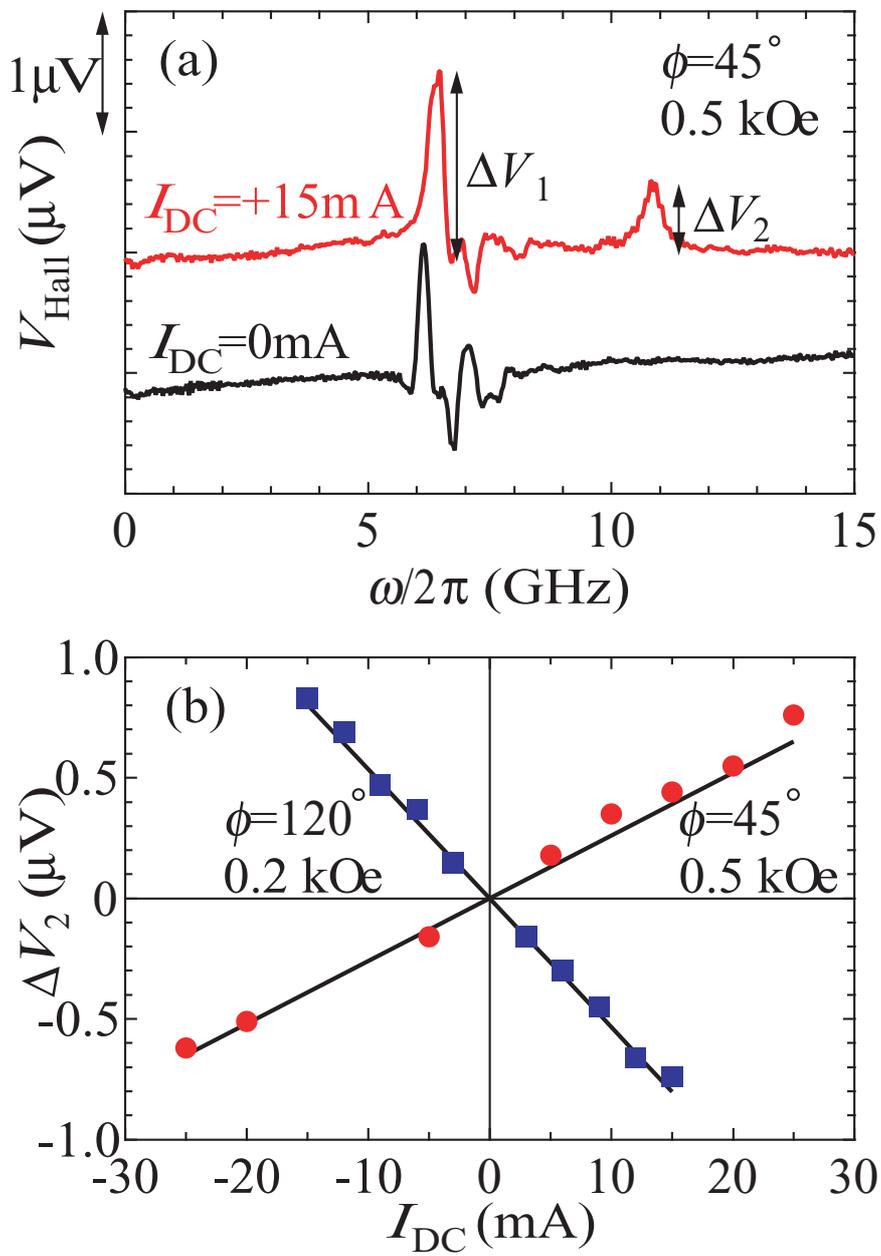

Fig. 2

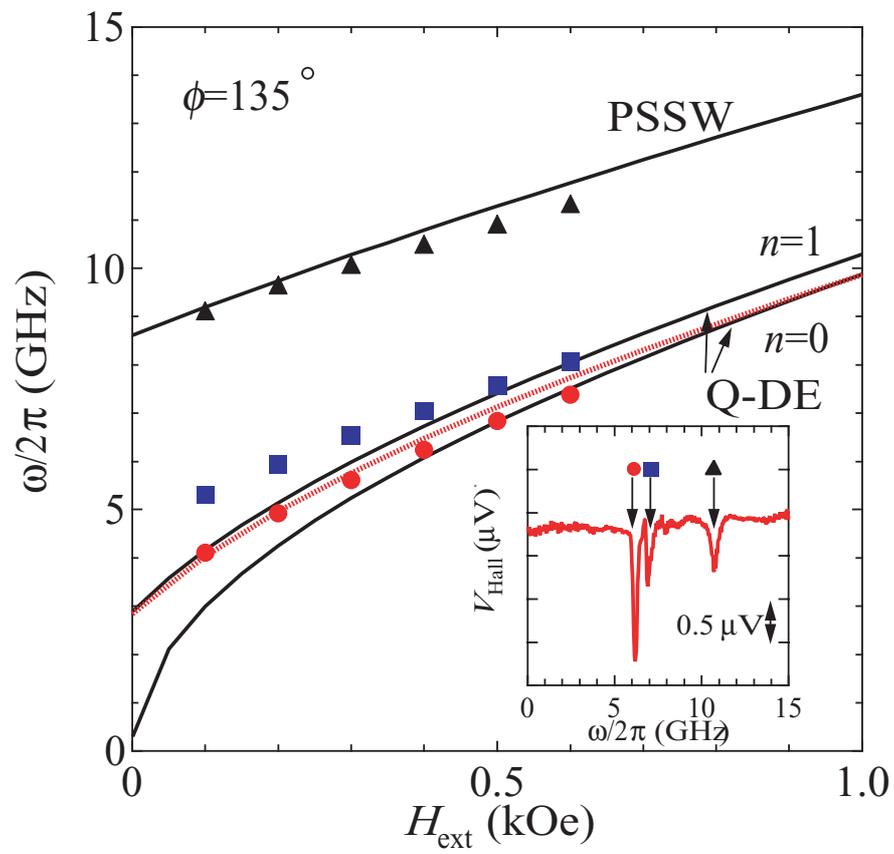

Fig. 3

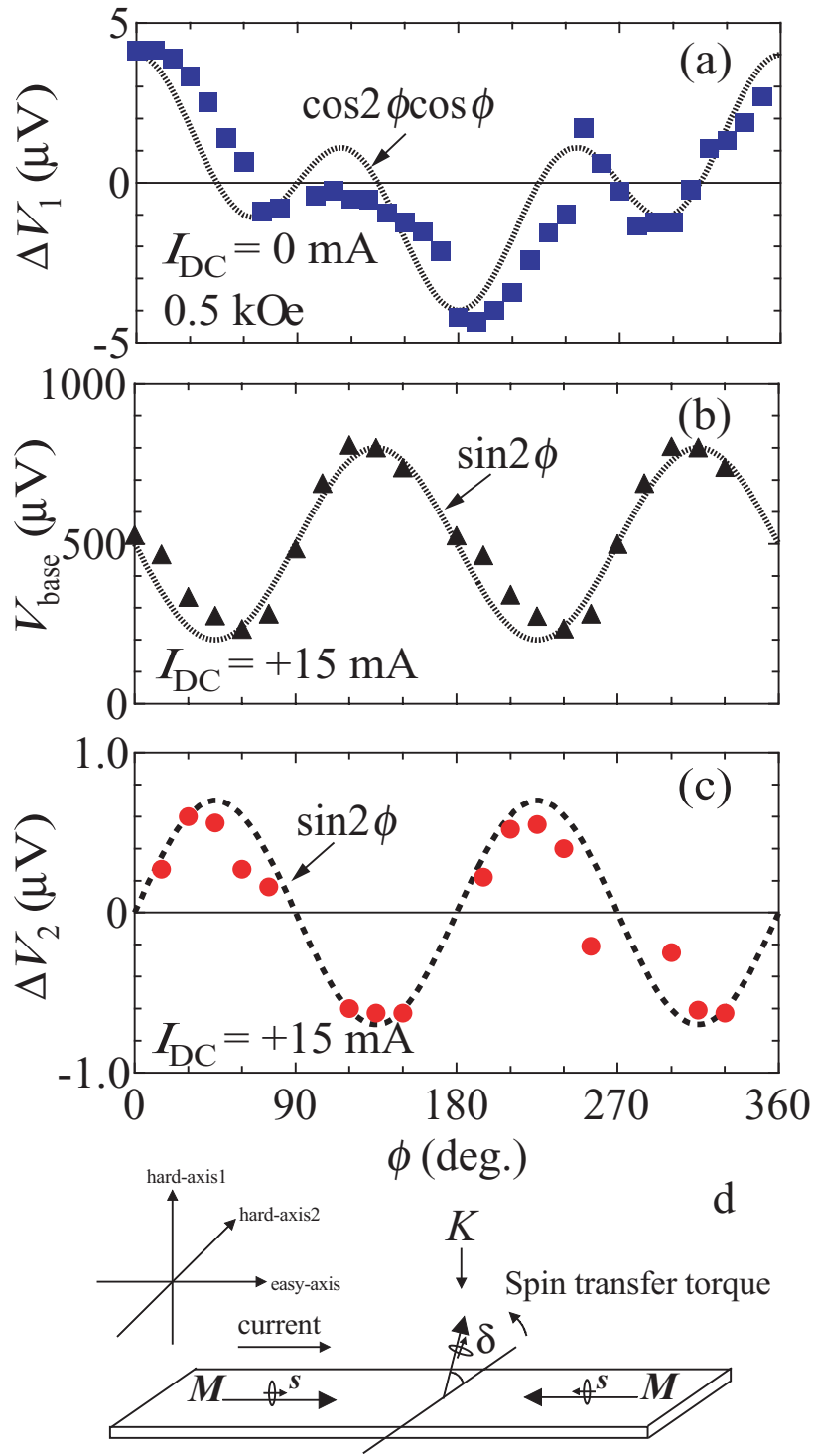

Fig. 4